\documentclass[aps,prd,twocolumn,superscriptaddress]{revtex4-1} 
\usepackage{fullpage, amsmath, xparse, amssymb, mathtools, graphicx, braket,xcolor,verbatim}
\usepackage{natbib}
\usepackage[colorlinks=true,linkcolor=blue,citecolor=blue]{hyperref}

\usepackage{multirow}
\usepackage{cancel}

\begin{document}
\title{Stochastic lensing of stars by ultralight dark matter halos}
\author{Andrew Eberhardt}
\thanks{Kavli IPMU Fellow. Corresponding author}
\email{\\ andrew.eberhardt@ipmu.jp}
\author{Elisa G. M. Ferreira}
\affiliation{Kavli Institute for the Physics and Mathematics of the Universe (WPI), UTIAS, The University of Tokyo, Chiba 277-8583, Japan}
\author{Wentao Luo}
\affiliation{Institute of Deep Space Sciences, Deep Space Exploration Laboratory, Hefei, Anhui 230088, People`s Republic of China}
\author{Shurui Lin}
\affiliation{Department of Astronomy, University of Illinois at Urbana-Champaign, 1002 West Green Street, Urbana, IL 61801, USA}
\author{Yin Li}
\affiliation{1 Department of Mathematics and Theory, Peng Cheng Laboratory, Shenzhen, Guangdong 518066, People`s Republic of China}

\begin{abstract}
Ultralight dark matter is an interesting dark matter candidate describing the lightest end of the mass parameter space. This model produces an oscillating granular pattern in halo densities. These fluctuations have the potential to produce a time-varying density along the line of sight creating a small lensing signal for any stars observed through a dark matter halo which oscillates on the de Broglie timescale. In this work, we study this stochastic lensing signal taking into account the impact of density granules as well as the central soliton. We calculate the amplitude and temporal properties of this signal and estimate how stellar observations may be used to constrain the ultralight dark matter mass and abundance. 
\end{abstract}

\maketitle

\section{Introduction}
The standard model of cosmology, $\Lambda$CDM, has proved very successful at describing large scale structure observations. However, the particle nature of the cold dark matter (CDM) component of this model remains a mystery. 
At the lowest mass end of the dark matter parameter space ($\lesssim 10^{-19} \, \mathrm{eV}$), ultralight (or fuzzy) dark matter (ULDM) is one interesting candidate \cite{Hu2000}.
The existence of many ultralight fields, one or more of which may be cosmic dark matter, is a generic prediction of many string theories \cite{Arvanitaki_2010}. However, ultralight dark matter was studied in cosmology in the context of small scale structure problems such as the core-cusp\cite{navarro1996, Persic1995, Gentile2004}, missing satellites \cite{Klypin1999, Moore1999}, and too-big-to-fail \cite{Boylan-Kolchin2011} problems (see \cite{Weinberg2015, Bullock2017} for review).

The wavelike phenomena of ultralight dark matter suppress structure below the scale of the de Broglie wavelength, $\lambda_\mathrm{db} = 2 \pi \hbar / m \sigma$, allowing a dark matter only solution to the absence of expected structure \cite{Hu2000}. However, currently, most small scale structure problems also admit observational or baryonic solutions and small scale structure observations are among the strongest constraints on the ultralight dark matter model. 
For example, current constraints include: the subhalo mass function ($> 3\times 10^{-21}\, \mathrm{eV}$) \cite{Nadler_2021, Schutz2020},
ultra-faint dwarf half-light radii ($> 3 \times 10^{-19}\, \mathrm{eV}$ \cite{Dalal2022, Marsh:2018zyw}) or cores~\cite{Hayashi:2021xxu}, 
galactic density profiles ($> 10^{-20}\, \mathrm{eV}$ \cite{Bar_2018, Bar_2022}),
satellite masses ($>6 \times 10^{-22}\, \mathrm{eV}$ \cite{Safarzadeh_2020}), 
Lyman-alpha forest ($>2 \times 10^{-20}\, \mathrm{eV}$ \cite{Rogers_2021}), 
strong lensing ($>4 \times 10^{-21}\, \mathrm{eV}$ \cite{Powell:2023jns}). For recent reviews, see \cite{Hui_2017, Ferreira_2021,Hui:2021tkt,NIEMEYER2020103787}.

When modeled as a single classical field, ultralight dark matter creates $\sim \mathcal{O}(1)$ oscillations in the density field that result in a granular structure in dark matter halo densities \cite{Schive2014_prl}. These density granules oscillate on the de Broglie time and length scale, $\tau_\mathrm{db} = 2 \pi \hbar / m \sigma^2$ and $\lambda_\mathrm{db}$, respectively. 
These oscillations have so far provided some of the highest mass constraints on this model \cite{Dalal2022, Marsh:2018zyw}. Interestingly, at masses $\gtrsim 10^{-17}\, \mathrm{eV}$, the oscillation timescales are similar or less than experimental observation times, i.e., $\sim 30 \, \mathrm{yrs}$, making astrophysical observations an exciting probe of ultralight dark matter which may test some of the higher masses. 
Already the impact of these granules on observations has been studied in the context of dynamic heating of stellar dispersions \cite{Marsh:2018zyw, Dalal2022, DuttaChowdhury2023}, astrometry \cite{Kim:2024xcr, dror2024}, and pulsar timing arrays \cite{Kim2024, Eberhardt:2024ocm}. 

In this work, we study the impact of the oscillating density structure on the lensing of light when viewed through the galactic halo. The stochastic oscillation of ultralight dark matter provides a stochastic lensing signal for all sources viewed through an ultralight dark matter halo. We quantify the amplitude and temporal properties of this signal for Milky Way-like halos considering both the central soliton and oscillating density granules.
Characterizing these phenomena is interesting for a number of reasons. Variation on the timescale of stellar observations could probe some of the highest masses yet for ultralight dark matter, a highly unexplored mass range, with peak sensitivity around $\gtrsim 10^{-17}, \mathrm{eV}$.
Additionally, this effect is highly serendipitous with current stellar observation missions, as the brightness of every star observed for essentially any other science goal provides a potential constraint on ultralight dark matter. Since stochastic lensing is ubiquitous in the ULDM model and affects all observed sources, understanding it is essential for using it as a constraint on this model.

We organize the paper as follows. In Section \ref{sec:background}, we discuss the relevant ultralight dark matter and lensing background. In Section \ref{sec:simulations} we describe our simulations, and the results of those simulations in Section \ref{sec:results}. Section \ref{sec:discussion} contains a discussion of the implications and caveats of these results. Finally, we summarize and conclude in Section \ref{sec:conclusions}.

\section{Background} \label{sec:background}
\subsection{Ultralight dark matter}
\subsubsection{Schr\"odinger-Poisson equations}

The most well studied ultralight dark matter model treats the dark matter as a classical spin-0 field. We will assume that the field can be treated as a complex, non-relativistic, classical field for the calculations contained in this paper. Though extensions to higher spins \cite{Amin2022}, quantum corrections \cite{Eberhardt2023}, and multiple fields \cite{Gosenca2023} also exist. The equations of motion for our field, $\psi$, are the Schr\"odinger-Poisson equations given
\begin{align} \label{eqn:SP}
    \partial_t \psi(x,t) &= -i \left( \frac{-\hbar \nabla^2}{2m} + \frac{mV(x,t)}{\hbar} \right) \psi(x,t) \, , \\
    \nabla^2 V(x,t) &= 4 \pi G |\psi|^2 \, ,
\end{align}
where $m$ is the dark matter particle mass, and the normalization of $\psi(x,t)$ is chosen such that its square amplitude equals the spatial density, i.e., $|\psi(x,t)|^2 = \rho(x,t)$. Oscillations in the field are usually described by the de Broglie length and time scales which are given respectively
\begin{widetext}
\begin{align}
    \lambda_\mathrm{db} &= 2 \pi \hbar / m \sigma \sim 6 \times 10^{-6} \left( \frac{10^{-17} \, \mathrm{eV}}{m} \right)  \left( \frac{200 \, \mathrm{km/s}}{\sigma} \right) \, \mathrm{kpc} \, . \\
    \tau_\mathrm{db} &= 2 \pi \hbar / m \sigma^2 \sim 30 \left( \frac{10^{-17} \, \mathrm{eV}}{m} \right) \left( \frac{200 \, \mathrm{km/s}}{\sigma} \right)^{2} \, \mathrm{yrs} \,,  \label{eqn:deBroglieTime}
\end{align}
\end{widetext}
where $\sigma$ is some characteristic velocity scale in the system, in this work we will take it to be the galactic dark matter velocity dispersion. 

It will also be useful to consider the field from the eigenvalue perspective~\cite{Lin:2018whl,Zagorac:2021qxq,Zagorac:2022xic}. Assuming a spherical symmetry, the eigenbasis associated with the radial potential, $V(r)$, is given
\begin{align}\label{eqn:eigenval}
     \left( \frac{-\hbar^2 \nabla_r^2}{2m} + m V(r) + \frac{\hbar^2}{2m}\frac{l(l+1)}{r^2} \right) \phi_{ln}(r) = E_{ln} \phi_{ln} \,,
\end{align}
where $l$ and $n$ are the angular momentum and energy quantum numbers, respectively. An arbitrary field can be constructed on this basis as
\begin{align} 
    \psi(r, \theta, \varphi) = \sum_{l=0}^{l_{max}} \sum_{m = -l}^{l} \sum_{n=0}^{e_{max}} w_{nlm} \, Y^m_l(\theta, \varphi) \, \phi_{ln}(r) \, e^{-i \, \omega_{lmn}} \,,
\end{align}
where $Y^m_l(\theta, \varphi)$ are the appropriate spherical harmonics, and $\omega_{lmn}$ are random phases for the corresponding eigenvectors. 

\subsubsection{Solitons} \label{subsec:solitons}

\begin{figure}[!ht]
	\includegraphics[width = .47\textwidth]{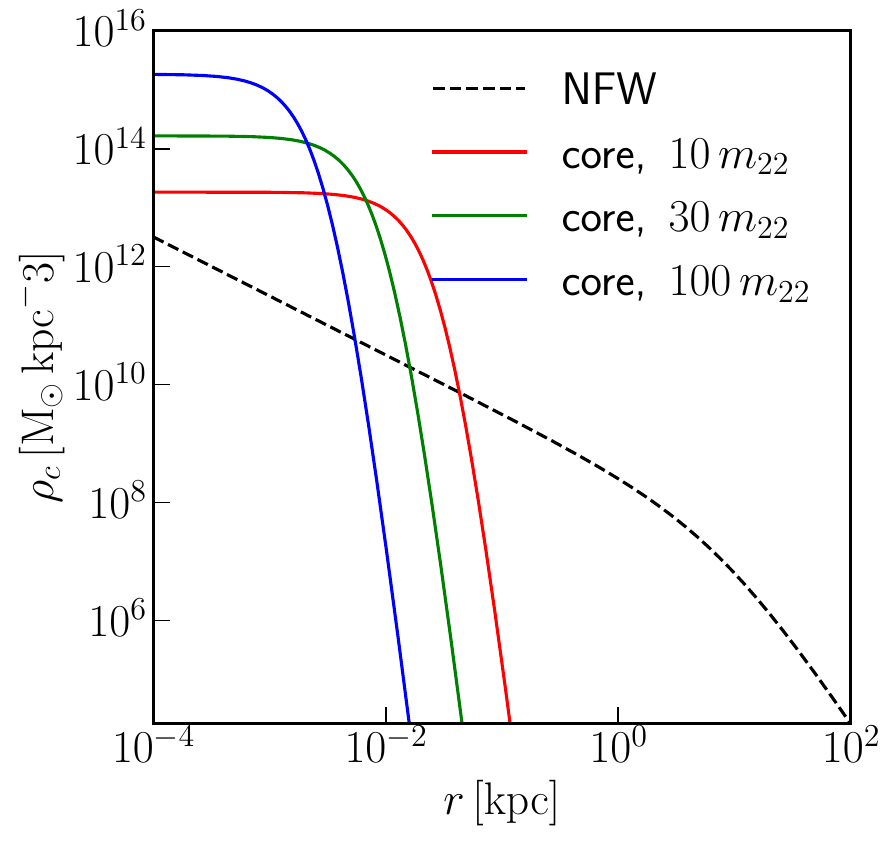}
	\caption{ The density of the NFW profile of the Milky Way \cite{Lin2019} and the core densities for three different mass FDM fields. }
	\label{fig:solitons}
\end{figure}
``Solitons" are the stable cores that form at the center of ultralight dark matter halos. The core profile is given by the ground state of the spherically averaged potential, i.e.,  $\phi_{00}$ in equation \eqref{eqn:eigenval}. The soliton profile is usually described as 
\begin{equation}
    \rho_\mathrm{core} (r) = |\phi_{00}|^2 = \frac{\rho_c}{[1+0.091 (r/r_c)^2]^8}\,,
\end{equation}
where $r_c$ is the soliton core radius and $\rho_c$ is the central density given by \cite{Schive2014_prl}
\begin{align}
    \rho_c &= 1.9 \left(  \frac{m}{10^{-23} \, \mathrm{eV}} \right)^{-2} \left(  \frac{r_c}{\mathrm{kpc}} \right)^{-4} \, \mathrm{M}_{\odot} \mathrm{pc}^{-3} \, , \\
    r_c &= 16 \,\left(  \frac{m}{10^{-23} \, \mathrm{eV}} \right)^{-1} a^{1/2} \left( \frac{M_\mathrm{vir}}{10^9 \mathrm{M}_\odot} \right)^{-1/3} \, \mathrm{kpc} \, .
\label{eq:Schive}
\end{align}
The radial density of a halo can usually be described as 
\begin{equation}
    \rho_\mathrm{halo} = \begin{cases}
    \rho_\mathrm{core} \,, \,\, r < r_\mathrm{tr} \\ 
    \rho_\mathrm{NFW} \,, \,\, r>r_\mathrm{tr} \,.
\end{cases}
\label{eq:rhohalo}
\end{equation}
for some transition radius $r_\mathrm{tr}$. We will assume here that the galactic profile well above the de Broglie scale is well described by an NFW profile
\begin{equation}
\rho_\mathrm{NFW} (r) = \frac{\rho_s}{(r/r_s) (1+r/r_s)^2}\,,
\end{equation}
with scale density, $\rho_s$, and scale radius, $r_s$. Soliton profiles are compared with an NFW profile for a few different ultralight dark matter field masses in Figure \ref{fig:solitons}. 

It was noted in simulations~\cite{Schive2020,Zagorac:2021qxq} that the soliton in the center of the galaxy is not stationary, but oscillates, performing a random walk around the halo center. The timescale associated with the motion of the soliton was determined to be
\begin{align} \label{eqn:tau_s}
    \tau_s = 120 \, \left( \frac{\rho_c}{0.1 \mathrm{M}_{\odot} \mathrm{pc}^{-3}} \right)^{-1/2} \, \mathrm{Myr} \, .
\end{align}

\subsubsection{Granules}

The higher energy and modes in our eigenvalue decomposition, i.e., \eqref{eqn:eigenval}, give us the granular density structure of the dark matter ``skirt". These correspond to the interference patterns present in ULDM given its wave nature. If we are far from the core, and look at scales small compared to the galactic scale radius we can approximate the field as a Gaussian random field with modes described by a Maxwell-Boltzmann distribution \cite{Hui_2021}, i.e.,
\begin{align}
    |\psi(k)|^2 \propto e^{-(1/2)(\hbar k / m \sigma)^2} \, .
\end{align}
The distribution of the density is given by 
\begin{align}
    P(\rho) = \frac{1}{\braket{\rho}} e^{-\rho/\braket{\rho}} \, ,
\end{align}
with correlation function of density fluctuations
\begin{align}
    \braket{\delta \rho(x) \, \delta \rho(x+r)} = \braket{\delta \rho^2} e^{- (r m \sigma / \hbar)^2} \, ,
\end{align}
where $\braket{\rho}$ is the average value of the density and $\braket{\delta \rho^2} = \braket{\rho}^2$ the variance. The correlations decay away exponentially beyond the scale of the de Broglie wavelength.

\subsection{Gravitational lensing}

We want to describe the lensing of light caused by the presence of substructures in the Galactic halo. Similar to what has been done to model the perturbation caused by the presence of the substructure of FDM in a lens ~\cite{Powell2023,Bartelmann2001}, we can write the lensing convergence (the projected surface mass density) for an observer at the origin and source a distance $D$ away along the line of sight as:
\begin{align}
    \kappa = \frac{4 \pi G}{c^2} \int_0^D dx \frac{x(D-x)}{D} \rho(x) \, , \label{eqn:lens}
\end{align}
where the $\rho(x)$ represents the substructure along the path. This expression can be used for the lensing caused by substructures present in the line of sight like the granules FDM or the soliton in the center of the halos.

\subsubsection{Granules}
\label{Sec:lens_granules}

In the case of the granules, the lensing convergence (\ref{eqn:lens}) is given by:
\begin{align}
    \delta \kappa_{\mathrm{gr}} = \frac{4 \pi G}{c^2} \int_0^D dx \frac{x(D-x)}{D} \delta \rho_\mathrm{gr}(x) \, , \label{eqn:lens_granules}
\end{align}
where the overdensity caused by the FDM substructure $\delta \rho_\mathrm{gr} = \rho_\mathrm{gr} - \rho_\mathrm{sm}$. We note that this describes the lensing on top of the smooth density profile $\rho_\mathrm{sm}$ and that $\delta \kappa_\mathrm{gr}$ can be positive or negative.

We approximate the dark matter granules as a Gaussian random field with correlation length equal to the de Broglie wavelength, $\lambda_\mathrm{db} = 2 \pi \hbar / m \sigma$, and expectation and variance given $\braket{\delta \rho} = 0$ and $\braket{\delta \rho^2} = \rho^2_{\mathrm{sm}}$, respectively (see \cite{Chan2020}). The expectation value, $\braket{\kappa_{\mathrm{gr}} }$ must be $0$ if we are to recover the correct smooth dark matter density. This field has variance
\begin{widetext}
\begin{align}
    \mathrm{Var}[\delta \kappa_{\mathrm{gr}}] &= \frac{16 \pi^2 G^2}{c^4} \int_0^D dx\, \int dx'\, \frac{x(x+x') (D - x) (D - x - x')}{D^2} \braket{\delta \rho(x) \, \delta \rho(x+x')} \, , \nonumber \\
    &= \frac{16 \pi^2 G^2}{c^4} \int_0^D dx\, \int dx'\, \frac{x(x+x') (D - x) (D - x - x')}{D^2} \braket{\delta \rho^2(x)} e^{-(2 \pi x'/\lambda_\mathrm{db})^2} \, , \nonumber \\
    &= \frac{8 \pi^{3/2} G^2}{c^4} \lambda_\mathrm{db} \int_0^D dx\, \frac{x^2 (D - x)^2 }{D^2} \braket{\delta \rho^2(x)} \, , \nonumber \\ 
    &= \frac{8 \pi^{3/2} G^2}{c^4} \lambda_\mathrm{db} \int_0^D dx\, \frac{x^2 (D - x)^2 }{D^2} \rho_\mathrm{sm}^2\, . \label{eqn:analyticGranules}
\end{align}
\end{widetext}
Here we added the distance factor to the integrand calculated in \cite{Chan2020}. We get from the second to the third line by assuming the distance factor changes a little over $\lambda_\mathrm{db}$.

The distribution of granules in a galactic halo can be well approximated by a sum of plane waves when we restrict ourselves to distances small compared to the galactic scale radii. In other words, this accurately describes the statistics of fluctuation when $D$ is small compared to the scale lengths of the galactic dark matter halo. If we make this approximation, then we can analytically integrate \eqref{eqn:analyticGranules}. In this case, $\rho_\mathrm{sm}(x) = \rho_0$, the average density, and all that remains is the integral over the distance factor, i.e.,
\begin{align}
    \mathrm{Var}[\delta \kappa_{\mathrm{gr}}] &= \frac{8 \pi^{3/2} G^2}{c^4} \lambda_\mathrm{db} \,  \rho_0^2 \int_0^D dx\, \frac{x^2 (D - x)^2 }{D^2} \nonumber \, , \\
    &= \frac{8 \pi^{3/2} G^2}{c^4} \lambda_\mathrm{db} \,  \rho_0^2 \frac{D^{3}}{30}\, . \label{eqn:kappa_rms_granules}
\end{align}
\begin{widetext}
    
To get a sense of the scaling of this signal, when the $D$ is small compared to the scale length, the root variance of $\kappa$ then scales as 
\begin{align} \label{eqn:lensing_granules}
    \delta \kappa_\mathrm{rms}^\mathrm{gr} \sim 3 \times 10^{-12} \left( \frac{10^{-17} \, \mathrm{eV}}{m} \right)^{1/2} \left( \frac{200 \, \mathrm{km/s}}{\sigma} \right)^{1/2} \left( \frac{\rho}{10^7 \, \mathrm{M_\odot / kpc^3}} \right) \left( \frac{D}{\mathrm{kpc}} \right)^{3/2} \, .
\end{align}
However, when our small distance approximation is not valid one must instead integrate equation \eqref{eqn:analyticGranules}.

\end{widetext}
\subsubsection{Soliton}
\label{Sec:lens_soliton}

The oscillations of the central soliton density can also provide a stochastic lensing signal. This oscillation is also $\sim \mathcal{O}(1)$, similar to the granular density pattern. We can approximate the amplitude of this oscillation using the core radius and central density found in \cite{Schive2014_prl}. Starting from equation \eqref{eqn:lens} we write the lensing from the soliton as
\begin{widetext}
\begin{align}
    \kappa_\mathrm{rms}^s &= \frac{4 \pi G}{c^2} \int_0^D dx \frac{x(D-x)}{D} \rho(x) \, \nonumber \\ 
    &\approx \frac{4 \pi G}{c^2} \frac{D_s (D-D_s)}{D} r_c \, \frac{\rho_c}{2} \, \\
    &\sim 10^{-8} \left( \frac{m}{10^{-23} \, \mathrm{eV}} \right)^{-2} \left( \frac{r_c}{\mathrm{kpc}} \right)^{-3} \left( \frac{D_s(D-D_s)/D}{\mathrm{kpc}} \right)
    \label{eq:lens_soliton}
\end{align}
Where we have assumed that $r_c \ll D\, ,D_s$. The core radius, central density, and time scale are given in Section \ref{subsec:solitons}.
\end{widetext}

\section{Simulations} \label{sec:simulations}

\subsection{Initial conditions}

In this section, we describe how we generate the initial conditions of our simulations. In general we will first define a spatial grid, $x_i$, giving the points at which out field will be defined, the spatial grid is given by
\begin{align}
    x_i = -L/2 + i \, dx  \, ,
\end{align}
and similarly for the $y$ and $z$ directions. Here, $i \in [0,N)$ is an integer and is the index for our cells, $L$ is the size of our simulation box, and $dx = L/N$ is the grid resolution.  

\subsubsection{Plane-waves} \label{sec:simulations:planewaves}

We use simple simulations of plane-wave superpositions in periodic boxes to study the impact of ultralight dark matter density granules on the stochastic lensing signal. We initialize our field as 
\begin{align}\label{eqn:psi_planewave}
    \psi(\vec r_{ijk}) = \sqrt{\rho_0} \sum_l^{n_s} e^{- i m \vec v_l \cdot \vec r_{ijk} / \hbar} \, ,
\end{align}
where $\vec r_{ijk} = (x_i, y_i, z_k)$, $n_s \gg 1$ is the number of plane waves used to generate the initial conditions. $\vec v_i$ is the velocity of the $i$-th plane wave, chosen isotropically and with amplitude drawn randomly from a Maxwell-Boltzmann distribution
\begin{align}
    f(v) = \sqrt{\frac{2}{\pi}} \frac{v^2}{\sigma^2} e^{-v^2 / 2 \sigma^2} \, . \label{eqn:maxwellian}
\end{align}
$\sigma$ is the velocity dispersion of the dark matter. We will use $\sigma = 200 \, \mathrm{km/s}$ in our plane-wave simulations. 

Note that each $v_l$ in equation \eqref{eqn:psi_planewave} must be chosen such that the field is continuous and differentiable on the boundaries. Therefore $mv_l / \hbar$ should be some integer multiple of $2\pi/L$.

\subsubsection{Halos}

We can construct dark matter halos using a superposition of eigenfunction solutions to equation \eqref{eqn:eigenval}. At a high level, the procedure for constructing a halo using this method is as follows
\begin{enumerate}
    \item Construct a target radial density, $\rho_t(r)$.
    \item Solve the eigenvalue problem associated with that target density's Hamiltonian, $H_t$.
    \item Weight radial eigenfunctions, $\phi_n^l(r)$, with weights, $w_n^l$ such that their squared sum reproduces the target density profile.
    \item Construct the field, $\psi(r,\theta,\phi)$, as a sum of the weighted eigenfunctions and spherical harmonics isotopically and with random phase.
\end{enumerate}
The above procedure will give isotropic halos.

The target radial density we will use is a cored NFW profile
\begin{align}
    \rho_{t}(r) = \left\{
\begin{array}{ll}
      \frac{(1.9\times 10^{7}) \, m_{22}^{-2} r_c^{-4}}{(1+0.091 \, (r/r_c)^2)^8}   \,,& r < r_c \\
      \frac{\rho_s}{\frac{r}{r_s} \left( 1 + \frac{r}{r_s} \right)^2} \,,& r \geq r_c \\
\end{array} 
\right. \, ,
\end{align}
in units of $\mathrm{M_\odot / kpc^3}$. Where $\rho_s$ is the scale density, $r_s$ the scale radius, $r_c$ the core radius, and the field mass is $m = m_{22} \times 10^{-22} \, \mathrm{eV}$.

This density profile then gives us a radial Hamiltonian
\begin{align} \label{eqn:}
    &H_l(r) = -\frac{\hbar^2}{2 m } \nabla_r^2 + m V_t(r) + \frac{\hbar^2}{2m} \frac{(l + 1)l}{r^2} \, , \nonumber \\
    &\nabla_r^2 V_t(r) = 4 \pi G \, \rho_t(r) \, .
\end{align}
Where $l$ is the angular momentum quantum number. The eigenvalue problem of this Hamiltonian can be solved numerically, this gives a spectrum of eigenfunctions, $\phi_n^l$,
\begin{align}
    H_l \, \phi_n^l = E_n^l \, \phi_n^l \, .
\end{align}
The $l$'s and $n$'s included in our halo construction are truncated at sufficiently large values, $l_\mathrm{max}$ and $e_\mathrm{max}$, respectively, so as to accurately describe the halo dynamics. We also note that we set the boundary conditions such that the halo is isolated (non-periodic boundary conditions).
The eigenvectors are then assigned weights, $w_n^l$, such that their sum approximates the target profile, i.e., 
\begin{align}
    \rho_t(r) &\approx \sum_{l=0}^{l_{max}} \sum_{m = -l}^{l} \sum_{n=0}^{e_{max}} |w_n^l|^2 \, |\phi_n^l(r)|^2 \, .
\end{align}
There are various scenes to choose the weights, we use an iterative optimizer available in the scipy package. Using these weights we construct the final field by summing over the eigenfunctions and appropriate spherical harmonics with random complex phase, i.e.,
\begin{align}
    \psi(r, \theta, \varphi) = \sum_{l=0}^{l_{max}} \sum_{m = -l}^{l} \sum_{n=0}^{e_{max}} w_n^l \, Y^m_l(\theta, \varphi) \, \phi_n^l(r) \, e^{-i \, \omega_{lmn}} \, .
\end{align}
$\omega_{lmn}$ is chosen uniformly and randomly from $[0,2 \pi)$. Note that the weight does not depend on $m$.

\subsection{Solver}

The field is evolved using the standard symplectic leap-frog integrator. This works by first solving the Schr\"odinger equation, i.e., equation \eqref{eqn:SP} as
\begin{widetext}
\begin{align}
    \psi(x,t + \delta t) =  \exp \left[ \int_t^{t + \Delta t} dt' \frac{-i}{\hbar} \left( \frac{-\hbar^2 \nabla^2}{2m} + m V(x,t') \right) dt' \right] \, \psi(x,t') \, . 
\end{align}
On intervals $\delta t$ sufficiently short compared to dynamical times  we can approximate this as 
\begin{align}
    \psi(x,t + \delta t) \approx \exp \left[ \frac{-i}{\hbar} \left( \frac{-\hbar^2 \nabla^2}{2m} + m V(x,t) \right) \delta t \right] \, \psi(x,t) \, . 
\end{align}
For a more detailed description of ``sufficiently short", see for example \cite{Garny2018,Eberhardt2020}. We can split the argument of the exponential into the product of two exponentials, each diagonal in a different basis as 
\begin{align}
    \psi(x,t + \delta t) \approx e^{i \delta t \, \hbar \nabla^2 / 2m} e^{-i \delta t \, m V(x,t) / \hbar} \left( 1+ \mathcal{O}(\delta t^2) \right) \, \psi(x,t) \, .
\end{align}
\end{widetext}
The left exponential is a ``drift" update and updates the position component of the ultralight dark matter phase space and is diagonal in the momentum basis. This portion of the update is 
\begin{align}
    \psi(t + \delta t) = \mathcal{F}^{-1}\left[ e^{i\hbar k^2 \delta t / 2m} 
    \mathcal{F}\left[ \psi_i(t) \right] \right] \, .
\end{align}
$\mathcal{F}$ is the Fourier transform.
The right exponential is a ``kick" update and updates the velocity component of the phase space and is diagonal in the position basis. 
\begin{align}
    \psi(t + \delta t) = e^{-i m V(x) \delta t / \hbar} \psi(t) \, . 
\end{align}
We can create a symplectic leap-fog kick-drift-kick scheme as follows
\begin{align} \label{eqn:pseudoSpecUpdate}
    \psi_1(x) &= e^{-i \delta t \, m V(x,t,\psi(t)) / \hbar / 2} \psi(x,t) \, , \\
    \psi_2(k) &= e^{i \delta t \frac{\hbar k^2}{2m} } \mathcal{F}\left[ \psi_1(x) \right](k) \, , \\
    \psi(x, t + \delta t) &= e^{-i \delta t \, m V(x,t,\psi_2) / \hbar / 2} \mathcal{F}^{-1}\left[ \psi_2(p) \right](x) \,,
\end{align}
where the potential is solved using the spectral method, i.e., 
\begin{align}
    V(x,t,\psi) = \mathcal{F}^{-1}\left[ 4 \pi G \,  \frac{\mathcal{F} [|\psi(x',t)|^2] (\vec k)}{k^2} \right](x) \, . \label{eqn:Poisson_k_space}
\end{align}

\section{Results} \label{sec:results}

Code used to make figures is available publicly at \href{https://github.com/andillio/Stochastic_Lens_FDM_public}{Stochastic\_Lens\_FDM\_public}. Simulation code and data can be made available upon request.

\subsection{Granules}

We use the lens model described in~\ref{Sec:lens_granules} to describe the stochastic lensing signal coming from the distribution of granules in the halo of the galaxy. 

\begin{figure*}[!ht]
	\includegraphics[width = .97\textwidth]{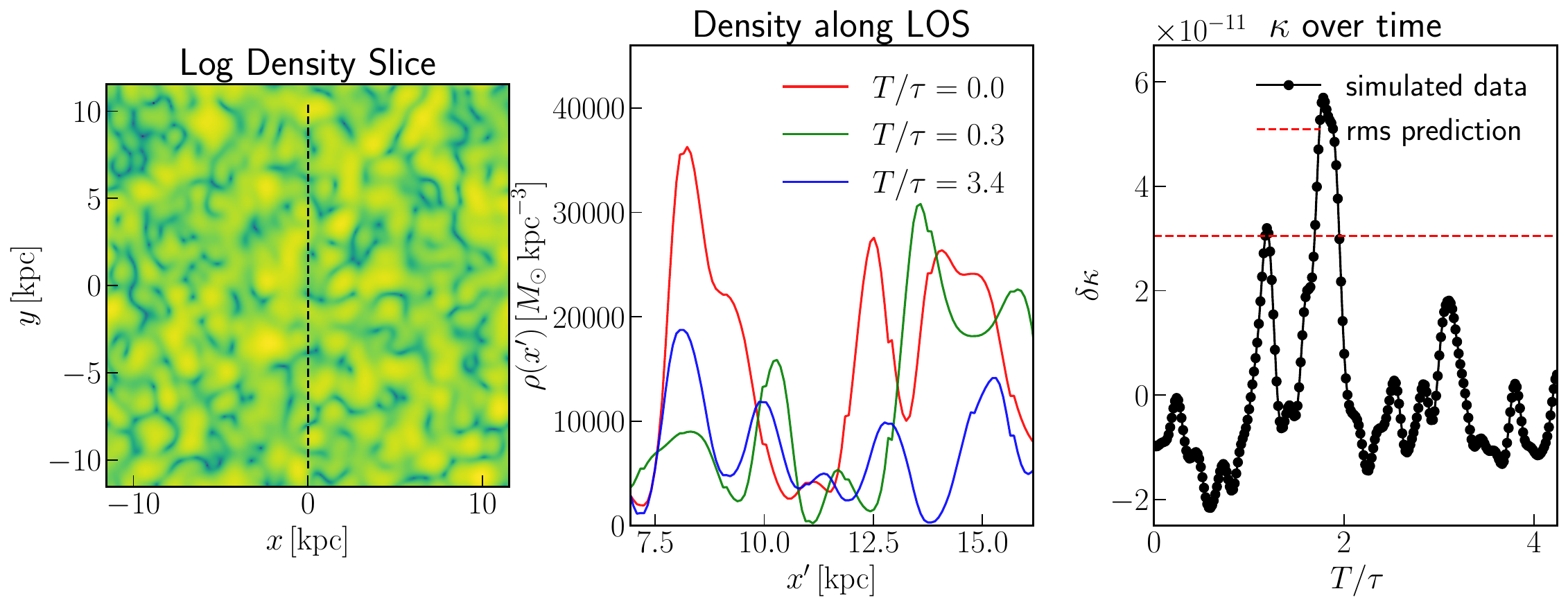}
	\caption{ Here we show the lensing due to the granules. \textbf{Left:} a density slice through an ultralight dark matter plane wave box. The black dotted line shows the line of sight through this slice. \textbf{Center:} the ultralight dark matter density along the line of sight at different times in the simulation showing the motion of the soliton. $T$ is the simulation time which we make dimensionless by dividing by the de Broglie time, $\tau$. \textbf{Right:} the integrated $\kappa$ value along this line of sight with mean subtracted. The dotted red line shows the predicted rms value which fits the measured rms value well. }
	\label{fig:granulesKappa}
\end{figure*}

\begin{figure}[!ht]
	\includegraphics[width = .44\textwidth]{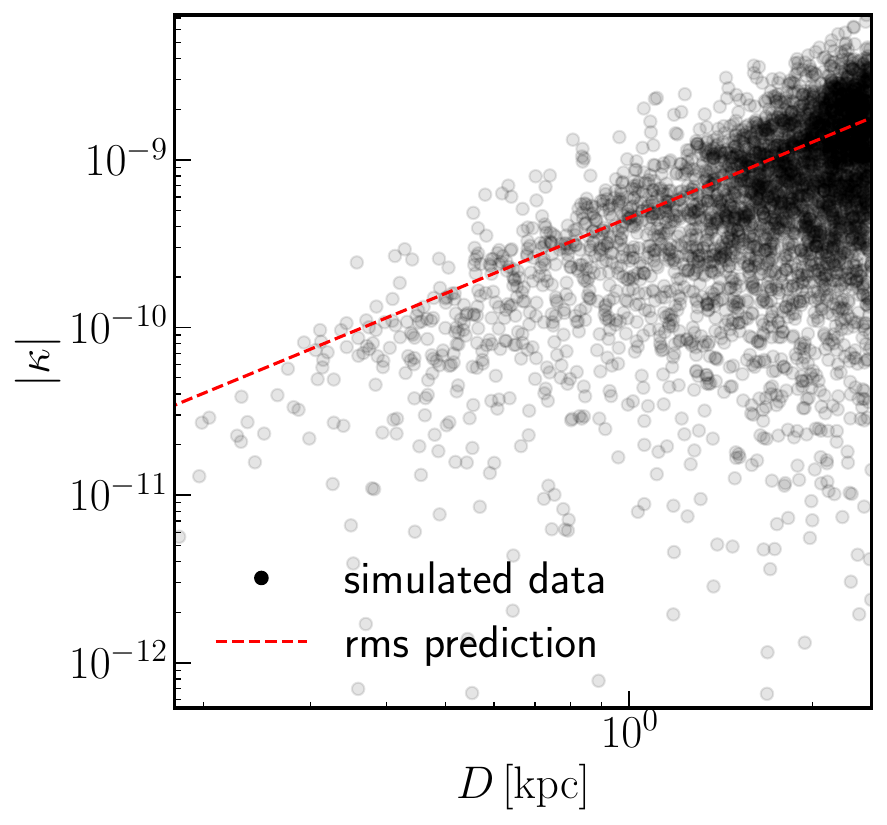}
	\caption{Comparison of stochastic lensing in plane wave simulation (see Section \ref{sec:simulations:planewaves}) and analytic calculation (i.e., equation \eqref{eqn:analyticGranules}). The black points represent the microlensing of sources randomly distributed in the box with an observer at the origin. The red line represents the $1\sigma$ prediction interval for the black points and should be larger than approximately $0.683$ of the points, and indeed in this simulation it contained $0.690$. In this simulation $m_{22} = 1$, $\sigma_\mathrm{dm} = 200 \, \mathrm{km/s}$, and $\rho_\mathrm{sm} = 0.01 \,  \mathrm{M_\odot /pc}^3$.  }
	\label{fig:granuleAnalytic}
\end{figure}

We study the effect of the granular stochastic lensing by simulating an oscillating ultralight dark matter density constructed of plane-waves super imposed in a periodic box. A density slice through one such simulation is shown in the left panel of Figure \ref{fig:granulesKappa}. We calculate the line integral of the density fluctuations along a hypothetical line of sight (black dashed line in left panel of Figure \ref{fig:granulesKappa}) and plot them at different simulation times, $T$, in the middle panel of Figure \ref{fig:granuleAnalytic}, the simulation time is made unitless by dividing out the de Broglie time, $\tau$. We track the lensing, $\delta \kappa$, caused by this density fluctuation as a function of time in the right panel of Figure \ref{fig:granuleAnalytic}. We plot the predicted rms fluctuation in dashed red to compare with the measured $\delta \kappa$ in the simulation. We can see that the amplitude of the fluctuation is of the same order as the prediction. 

We then track the magnification, $\delta \kappa$, for many points in a simulation in Figure \ref{fig:granuleAnalytic} and plot the corresponding $\delta \kappa$ value as a function of the distance from the observer (black points in Figure \ref{fig:granuleAnalytic}). We plot the rms prediction as a function of distance in dashed red. The dashed red line is above $0.69$ of the data points indicating that the rms prediction correctly provides a 1$\sigma$ prediction interval for the the granular stochastic lensing. 


\begin{figure}[!ht]
	\includegraphics[width = .44\textwidth]{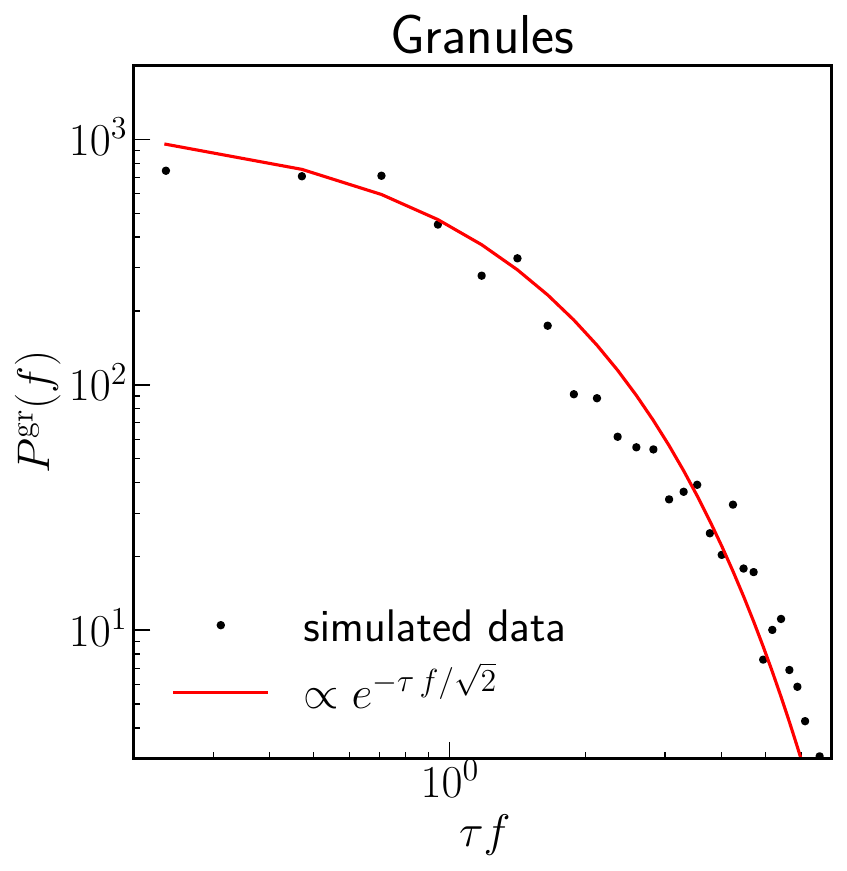}
	\caption{ The average temporal powerspectrum of $\kappa$ for a large number of points at $1 \, \mathrm{kpc}$ in a plane-wave box simulation with mass $15 m_{22}$.  }
	\label{fig:granulePS}
\end{figure}

We can also find the temporal power spectrum of the $\delta\kappa$ fluctuations. The power spectrum of the density is well described by 
\begin{align}
    P^\rho (k) \propto e^{-\hbar^2 k^2 / 2\sqrt{2} \sigma^2 m^2} \label{eqn:rhoPS} \, .
\end{align}
For a free field, we can relate spatial and temporal modes as 
\begin{align}
    f = E /2 \pi \hbar = \frac{\hbar^2 k^2}{2m} / 2 \pi \hbar = \frac{\hbar k^2}{4 \pi m} \, .
\end{align}
In frequency space then the temporal power spectrum describing the density fluctuations is simply
\begin{align} \label{eqn:PS_granules}
    P^\rho (f) \propto e^{-f \tau_\mathrm{db}} \, .
\end{align}
If we track the $\delta \kappa$ at $1 \, \mathrm{kpc}$ in our simulations as a function of time at a large number of points and then find the average power spectrum, we see that it is well-described by equation \eqref{eqn:PS_granules}. This is done in Figure \ref{fig:granulePS}. This power spectrum describes the time scales on which the signal evolves. We can see that the power in the signal is relatively constant for timescales longer than the de Broglie time but then exponentially decays for timescales shorter than the de Broglie time. This means realistically, to detect a signal an experiment would need to observe for at least the de Broglie timescale. The de Broglie time is inversely proportional to the mass, see equation \eqref{eqn:deBroglieTime}, and so sensitivity to lower masses requires longer observation times. Likewise, the amplitude of the fluctuation signal grows as the mass decreases, see equation \eqref{eqn:kappa_rms_granules}. This means that longer experiment integration times will in general allow access to lower masses and larger signals. 

\begin{figure}[!ht]
	\includegraphics[width = .44\textwidth]{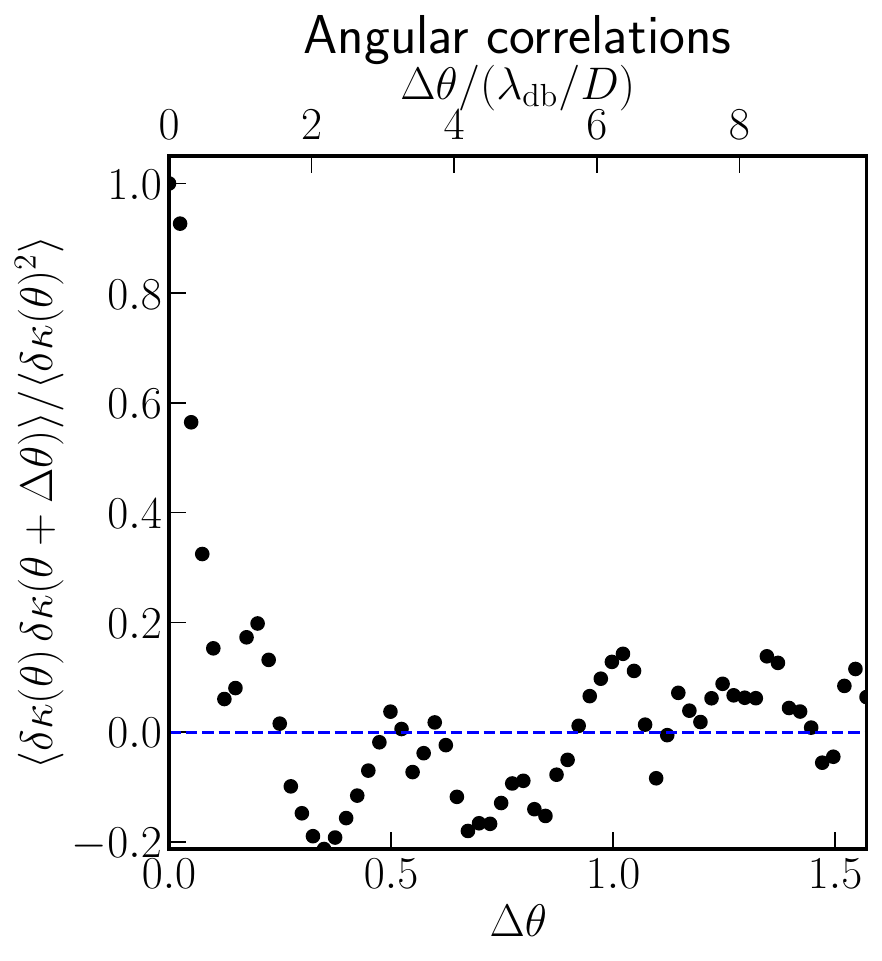}
	\caption{ Time averaged angular correlations in stochastic lensing signal. The correlation quickly decays and randomly fluctuates around $0$ on the scale of the de Broglie wavelength over the source-observer separation, $\lambda_\mathrm{db} / D$. This data is from a plane-wave box simulation run for $10 \, \mathrm{Gyr}$, $15 m_{22}$, $D = 10 \, \mathrm{kpc}$. }
	\label{fig:angularCorrelations}
\end{figure}

Figure \ref{fig:angularCorrelations} shows the time averaged angular correlation of the $\delta \kappa_\mathrm{gr}$ signal for the stochastic lensing from the granules. We can see that the correlation between stochastic lensing signals decays rapidly on angular separations greater than $\lambda_\mathrm{db} / D$ and then fluctuates around $0$. 

\subsection{Soliton}
\begin{figure*}[!ht]
	\includegraphics[width = .97\textwidth]{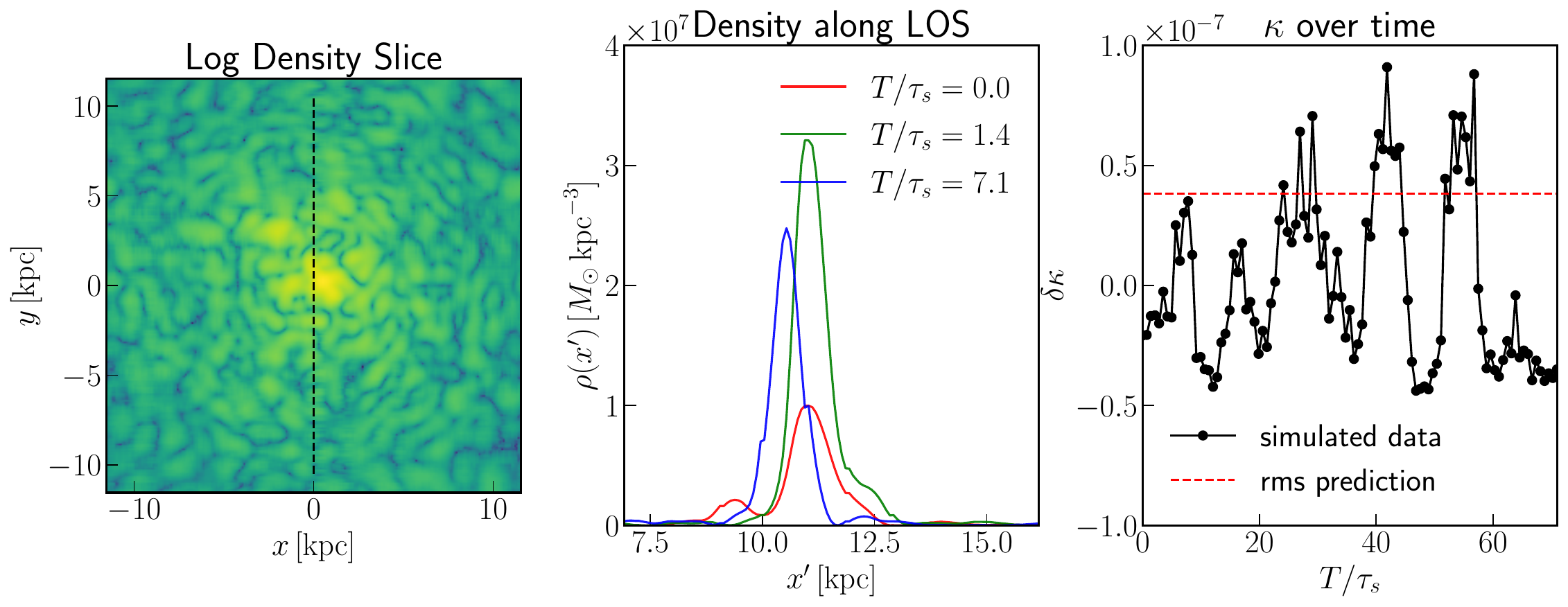}
	\caption{ Here we show the lensing due to the soliton. \textbf{Left:} a density slice through an ultralight dark matter halo. The black dotted line shows the line of sight through this slice. \textbf{Center:} the ultralight dark matter density along the line of sight at different times in the simulation showing the motion of the soliton. \textbf{Right:} the integrated $\kappa$ value along this line of sight with mean subtracted. The dotted red line shows the predicted rms value which fits the measured rms value well.}
	\label{fig:solitonKappa}
\end{figure*}

Modeling the lensing caused by the oscillation of the soliton as in Sec.~\ref{Sec:lens_soliton}, we compare the theoretical prediction in (equation \eqref{eq:lens_soliton}) with results of a halo simulation, shown in Figure \ref{fig:solitonKappa}. The left panel of Figure \ref{fig:solitonKappa} shows a slice through the simulated ultralight dark matter halo with a hypothetical line of sight (dashed black). The middle panel shows the density along this line of sight at different simulation times, $T$, made dimensionless by dividing by the time of the soliton oscillation timescale (equation \eqref{eqn:tau_s}). The right panel shows the lensing caused by the oscillation of the soliton over time (black line) compared with the rms prediction (red dashed line). We can see that the amplitude and timescale of the oscillation corroborate our predictions. 

The soliton has a similar size and oscillation timescale compared with the granules, however it is much more massive. Therefore, the amplitude of the lensing is large compared to that of the granules. Observationally, though using the soliton as a lens is somewhat challenging because there is only one and it is located in the center of the halo, whereas, in contrast, the granules are everywhere and tightly packed throughout the halo.

Notice that the scaling of $r_c$ with the mass in equation \eqref{eq:Schive} implies that $\kappa_s$ actually scales linearly with the ultralight dark matter mass. The caveat is of course that at higher masses the central core becomes very small. 

\section{Discussion} \label{sec:discussion}

\begin{figure}[!ht]
	\includegraphics[width = .47\textwidth]{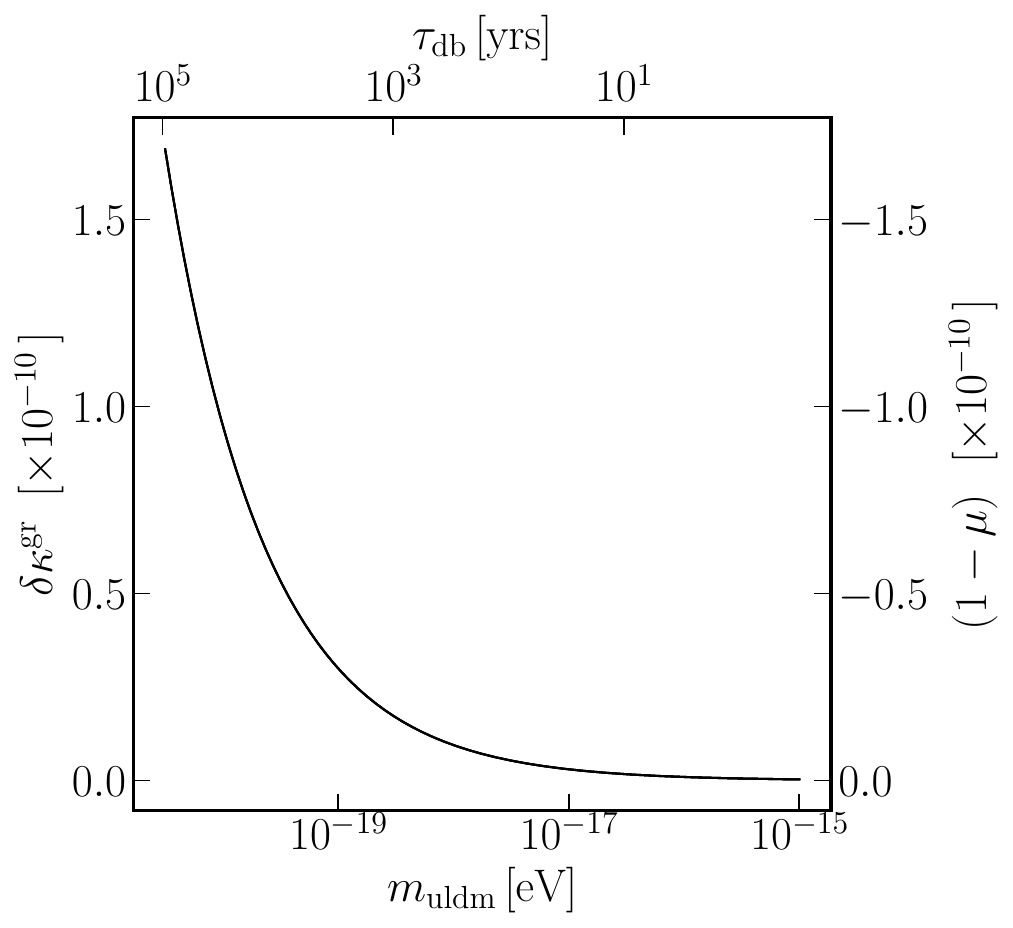}
	\caption{Stochastic lensing signal from the granules with respect to the ULDM mass and de Broglie time. We plot the lensing convergence and also the corresponding magnitude change in the source caused by the lensing.}
	\label{fig:mass_time_kappa_mag}
\end{figure}

In general, for granular lensing, the signal is going to be stronger when viewed through a larger portion of dark matter and for lower masses. To maximize the former we want to look at quiet and bright sources far away. To maximize the latter we need longer experimental observation times. In addition to the granule size, the dark matter mass also sets the timescales on which the signal fluctuates. In order to see this fluctuation we need to observe for at least the de Broglie time. This means that longer observations allow us to observe lower masses which have longer de Broglie time and larger fluctuation amplitudes. This interplay between the size of the stochastic lensing signal for each mass and its correspondent de Broglie time is summarized in Fig.~\ref{fig:mass_time_kappa_mag}. In this figure, we can see for the case of the granules, how $\delta \kappa^{\mathrm{gr}}_{\mathrm{rms}}$ changes for each mass of the ULDM particle and the correspondent de Broglie time scale. We also plot the signal with respect to the (bolometric) magnitude change caused by the stochastic lensing ($\delta m$)~\cite{Alsing:2014fya}.

One major potential caveat for this work is the potential presence of ultralight dark matter substructure at high masses, for example, miniclusters \cite{Ellis_2022}. When calculating the lensing due to granules we assumed that the ultralight dark matter was approximately smoothly distributed between us and the lensed star, if a substantial substructure exists then this calculation would need to be altered \cite{Ellis_2022}. 

The $\kappa$ values resulting from ultralight dark matter fluctuations are small compared to the magnifications induced by line of sight crossing of compact substructures like primordial black holes \cite{Niikura2019}. However, if ultralight dark fluctuations exist in every ultralight dark matter density. This means that there should be a background fluctuation in the brightness of every luminous object viewed through an ultralight dark matter halo and that these fluctuations should be correlated on the de Broglie scale (see Figure \ref{fig:angularCorrelations}). The angular correlations of the signal mean that tight groupings of stars (for example globular clusters \cite{Harris2010}) may be one of the best candidates for this method. This would have the benefit of allowing an averaging over the constituent stellar fluctuations and producing an overall brighter signal than observations of a single star. 

This technique of constraining ultralight dark matter is potentially very serendipitous with exoplanet transit surveys which monitor a large number of stars for luminosity fluctuations \cite{Gilliland_2011, Rauer2014, Ricker2016}. It should be possible to look for the fluctuations on the temporal scales in Figure \ref{fig:granulePS} in this kind of data. However, the current projected sensitivity or integration time of these surveys is much smaller than would be required to detect the fluctuations we discuss in this paper. This nonetheless is a potentially interesting possible use of constantly improving observational instruments and data. 

Strong lensing observations offer another potentially interesting source \cite{Khullar2021}. This probe has already been used to study ultralight dark matter \cite{Powell2023,Laroche:2022pjm,Chan:2020exg}, namely constraining these models by testing the presence of granules in the lens. In these works, it was pointed out that the effect of line of sight substructure can also be imprinted in the strong lens signal and needs to be taken into account for a precise understanding of the signal.  These works worked in a different mass range than the one considered here, for smaller masses, but show the potential of observable to test stochastic lensing. In our case, strong lensing is beneficial for a number of reasons. First, like in the globular cluster case, strongly lensed sources allow us to potentially average over many fluctuating objects. Second, these sources are viewed through the entire Milky Way dark matter halo, maximizing the number of granules that the signal passes through and increasing the lensing. 



\section{Conclusion} \label{sec:conclusions}
In this work, we studied the lensing arising from the stochastic fluctuations of ultralight dark matter halos. We studied both the lensing arising from the granular density fluctuations in the halo ``skirt" and that arising from the oscillations of the central soliton numerically and analytically. 
We find that the stochastic lensing signal for the density granules and central soliton go, respectively, as 
\begin{widetext}
\begin{align}
    \delta \kappa_\mathrm{rms}^\mathrm{gr} &\sim 3 \times 10^{-12} \left( \frac{m}{10^{-17} \, \mathrm{eV}} \right)^{-1/2} \left( \frac{\sigma}{200 \, \mathrm{km/s}} \right)^{-1/2} \left( \frac{\rho}{10^7 \, \mathrm{M_\odot / kpc^3}} \right) \left( \frac{D}{\mathrm{kpc}} \right)^{3/4} \, , \nonumber \\
    \kappa_\mathrm{rms}^\mathrm{s}    &\sim 10^{-8} \left( \frac{m}{10^{-23} \, \mathrm{eV}} \right)^{-2} \left( \frac{r_c}{\mathrm{kpc}} \right)^{-3} \left( \frac{D_s(D-D_s)/D}{\mathrm{kpc}} \right) \nonumber \, .
\end{align}
\end{widetext}
While the amplitude of the signal is small, particularly at masses where the de Broglie time is of order experimental observation times, importantly, every star or observation viewed through a dark matter halo provides a potential constraint. 
Because of the de Broglie scale temporal and spatial fluctuations, the lensing signal we identify here has a spectrum unique to ultralight dark matter making it straightforward to distinguish from other potential lensing sources. Likewise, the intrinsic variation of lenses sources can be distinguished either by looking for spatial correlations in the signal or by looking to see if the timescale of the fluctuation has the predicted power spectrum. Investigating the potential of various observational methods to constrain ultralight dark matter (ULDM) using this effect is currently a work in progress.
An important caveat is that this work was done assuming that the granules were smoothly distributed between observer and lensed star, we expect the calculation to change if ultralight dark matter substructure exists on the scale of the observer-star separation. We leave this as potential future work.

\bibliography{BIB}

\end{document}